%% file: mnras_template.tex
\documentclass[fleqn,usenatbib]{mnras}
\usepackage{newtxtext,newtxmath}
\usepackage[T1]{fontenc}

\DeclareRobustCommand{\VAN}[3]{#2}
\let\VANthebibliography\thebibliography
\def\thebibliography{\DeclareRobustCommand{\VAN}[3]{##3}\VANthebibliography}

\usepackage{graphicx}	% Including figure files
\usepackage{amsmath}	% Advanced maths commands
\usepackage{orcidlink}
\usepackage[english]{babel}
\usepackage[final]{changes}
\newcommand{\EBV}{$E(B-V)$}
\newcommand{\metal}{$12 + \text{log\,(O/H)}$}
\newcommand{\logq}{$\text{log}\,q$}
\newcommand{\Hb}{H ${\beta}$}

\newcommand{\tauint}{$\tau_{\text{int}}$}

\newcommand{\Ha}{H ${\alpha}$}

\newcommand{\SoS}{[\ion{S}{III}]/[\ion{S}{II}]}
\newcommand{\SoSinv}{[\ion{S}{II}]/[\ion{S}{III}]}
\newcommand{\SoSfull}{[S III]$\lambda\lambda9068,9532$ / [S II]$\lambda6717,6731$}
\newcommand{\logep}{$\text{log}\,\epsilon$}
\newcommand{\mhund}{`mock\_snr100'}
\newcommand{\mtena}{`mock\_snr010\_a'}
\newcommand{\mtenb}{`mock\_snr010\_b'}
\newcommand{\mtenc}{`mock\_snr010\_c'}
\newcommand{\mtend}{`mock\_snr010\_d'}
\newcommand{\mtene}{`mock\_snr010\_e'}

\title[Parameter Recovery Study on IZI]{Parameter Recovery Study on  IZI---a Bayesian Analysis Tool for Emission Lines from \ion{H}{II} Regions and Star-forming Galaxies}

\author[J.-H. Shinn et al.]{
Jong-Ho Shinn,$^{1 \orcidlink{0000-0001-7967-6473}}$\thanks{E-mail: jhshinn@kasi.re.kr}
Rory Smith,$^{2,3 \orcidlink{0000-0001-5303-6830}}$
and Kyuseok Oh$^{1 \orcidlink{0000-0002-5037-951X}}$
\\
$^{1}$Korea Astronomy and Space Science Institute, 776 Daeduk-daero, Yuseong-gu, Daejeon, 34055, the Republic of Korea\\
$^{2}$Departamento de F\'isica, Universidad T\'ecnica Federico Santa Mar\'ia, Avenida Espa\~na 1680, Valpara\'iso, Chile\\
$^{3}$Millennium Nucleus for Galaxies (MINGAL)
}

\date{Accepted XXX. Received YYY; in original form ZZZ}

\pubyear{\the\year{}}

\begin{document}
\label{firstpage}
\pagerange{\pageref{firstpage}--\pageref{lastpage}}
\maketitle

\begin{abstract}
We present a series of parameter recovery test results of the Bayesian analysis tool IZI, which analyses emission lines from \ion{H}{II} regions and star-forming galaxies and returns the estimates of the gas metallicity \metal{}, ionisation parameter \logq{}, and nebular emission-line colour excess \EBV{}.
We created several mock datasets using IZI to represent a few different ideal or realistic datasets and performed parameter estimation on the mock data with IZI.
We found that IZI underestimated or overestimated the parameters by approximately 1-$\sigma$ or greater when the model error was included, even when using all emission lines available in the model grids.
We strongly recommend that IZI users run parameter recovery tests adjusted for their data before interpreting the IZI estimates.
To encourage the appropriate use of IZI, we also share a script for parameter recovery tests.
The cause of IZI's biased estimation is the substantial model error in the likelihood term, which varies with the model parameters.
We thus note that any parameter estimation with a substantial, varying model error in the likelihood term could return biased estimates for the model parameters, such as in the case of NebulaBayes, another Bayesian analysis tool for photoionisation emission lines.
We also note two issues relevant to setting the \logq{} prior using the observed line ratio \SoSfull{} (the mismatch of line flux terms and violation of Bayes' theorem) and propose a way to avoid the issues.
\end{abstract}


%%%%%%%%%%%%%%%%%%%%%%%%%%%%%%%%%%%%%%%%%%%%%%%%%%

%%%%%%%%%%%%%%%%% BODY OF PAPER %%%%%%%%%%%%%%%%%%

\section{Introduction}
% galaxy evolution, star formation, H II region, emission lines
Galaxies evolve in diverse manners since their formation over the cosmic time.
They merge \citep{Conselice_2003_AJ_126_1183,Tacconi_2008_ApJ_680_246,Lotz_2011_ApJ_742_103}, change their morphologies \citep{Brinchmann_1998_ApJ_499_112,Lotz_2008_ApJ_672_177,Conselice_2014_ARA&A_52_291} and colours \citep{Bell_2004_ApJ_608_752,Whitaker_2011_ApJ_735_86}, accrete nearby gas \citep{Block_2002_A&A_394_L35,Sancisi_2008_A&ARv_15_189,Papovich_2011_MNRAS_412_1123}, form new stars of diverse masses \citep{Kennicutt_1998_ARA&A_36_189,Madau_2014_ARA&A_52_415}, and enrich themselves with newly synthesised elements from supernovae \citep{Nomoto_2013_ARA&A_51_457}.
Among these processes, star formation is of special interest, because it changes the properties of galaxies in various ways.
Star formation consumes gas and dust, which are the raw materials for new stars \citep{McKee_2007_ARA&A_45_565,Zinnecker_2007_ARA&A_45_481,Klessen_2023_ARA&A_61_65}; adds new stars, which changes the galaxy's spectrum \citep{Conroy_2013_ARA&A_51_393}; and generates ultraviolet radiation and stellar winds, which ionise and transform surrounding environments \citep{Yorke_1986_ARA&A_24_49,Shields_1990_ARA&A_28_525,Puls_2008_A&ARv_16_209,Eldridge_2022_ARA&A_60_455}.
Recently formed massive stars ($\gtrsim8\,M_\odot$, \citealt{Zinnecker_2007_ARA&A_45_481}) create ionised regions around themselves---\ion{H}{II} Regions---where diverse emission lines emanate \citep{Osterbrock_2005_book}.
These emission lines enable astronomers to extract the host galaxy's properties, such as the gas metallicity, and multiple methods have been developed for gas metallicity estimation.
These methods use the electron temperature, metal recombination lines, photoionisation model, and strong emission lines \added{\citep{Peimbert_2017_PASP_129_82001,Maiolino_2019_A&ARv_27_3,Aller_1954_ApJ_120_401,Peimbert_1967_ApJ_150_825,Kewley_2002_ApJS_142_35}}. 

% analysis using photoionisation model, IZI, IZI-citing studies
Among such methods, photoionisation models have several advantages in gas metallicity estimations \citep{Maiolino_2019_A&ARv_27_3}.
It has no limits on the parameter range that can be explored in principle and provides information about the ionisation property in addition to the metallicity.
It enables calibrating strong emission lines, where the direct electron-temperature method is unfavourable.
IZI, coined from `I'nferring the gas metallicity (`Z') and `I'onization parameter, is an analysis tool that uses photoionisation model grids \citep{Blanc_2015_ApJ_798_99,Mingozzi_2020_A&A_636_A42}.
IZI analyses the emission line ratios and returns Bayesian estimates of the gas metallicity, ionisation parameter, and nebular emission-line colour excess.
Numerous studies have used IZI for their analyses, and their topics are diverse---for example, the impact of diffuse ionised gas on the emission-line ratios from star-forming galaxies \citep{Zhang_2017_MNRAS_466_3217}, evolution of the dust-to-metal ratios in galaxies \citep{DeVis_2019_A&A_623_A5}, metallicity variations across galactic disks \citep{Kreckel_2019_ApJ_887_80}, host galaxy of a fast radio burst source \citep{Xu_2022_Nature_609_685}, galactic azimuthal variations of oxygen abundance \citep{Ho_2017_ApJ_846_39}, mass-metallicity relation in galaxies \citep{BarreraBallesteros_2017_ApJ_844_80}, value-added catalogue of galactic spectral energy distributions \citep{Chilingarian_2017_ApJS_228_14}, and \ion{Ly}{$\alpha$} emitter at the epoch of reionisation \citep{Jung_2024_ApJ_967_73}.

% problems reported for strong emission line analysis by myself, model error in likelihood
Meanwhile, \cite{Shinn_2020_MNRAS_499_1073} reported a possible overestimation of model parameters in studying the gas metallicity estimation of star-forming galaxies.
\cite{Shinn_2020_MNRAS_499_1073} modelled emission line fluxes based on the relationships between the metallicity and line-flux ratios, which is used in the strong line method described above \citep{Peimbert_2017_PASP_129_82001,Maiolino_2019_A&ARv_27_3}.
\cite{Shinn_2020_MNRAS_499_1073} found that the nebula emission-line colour excess and intrinsic emission-line fluxes can be overestimated by up to a factor of five and showed that the overestimation stems from the model error in the likelihood term, which varies with model parameter.
Considering this, we noticed that IZI includes the same type of model error in its likelihood term.
This naturally led us to investigate IZI further, since the model error might cause an underestimation or overestimation of the model parameters, in which case we aim to guide the community on the proper use of IZI.

% what I've done here, test of IZI
Here, we present a series of results on a parameter recovery test of IZI.
We created several mock datasets using IZI that represent different ideal or realistic cases and performed parameter estimation on the mock data using IZI.
IZI should recover the values of the input model parameter used to create the mock data if it does not suffer from any estimation bias; however, we found that one parameter was underestimated or overestimated by more than 1-$\sigma$ for the mock datasets mimicking realistic observations when the model error was included in the likelihood term.

\section{IZI: Bayesian Analysis Tool for Emission Lines from H II Regions and Star-forming Galaxies} \label{sec:izi}
% model parameter: gas metallicity, ionisation parameter, color excess

% what is IZI?
IZI is a Bayesian analysis tool for emission lines from \ion{H}{II} regions and star-forming galaxies.
It was originally written in IDL by \cite{Blanc_2015_ApJ_798_99} and designed to return the estimates of gas metallicity and ionisation parameter from the Bayesian analysis of the emission line fluxes from observations and photoionisation model grids.
Subsequently, \cite{Mingozzi_2020_A&A_636_A42} modified IZI to include a third parameter---nebula emission-line colour excess (\EBV{})---porting the code to Python.
This modification enabled \EBV{} inferencing using IZI considering the nebula temperature variation, which is generally neglected when deriving \EBV{} using the Balmer decrement, by simultaneously determining the gas metallicity, ionisation parameter, and colour excess \citep{Mingozzi_2020_A&A_636_A42}.
The gas metallicity estimate is given by the oxygen abundance \metal{} \citep{Maiolino_2019_A&ARv_27_3}, and the ionisation parameter estimate is given by \logq{}, where the ionisation parameter $q$ is defined as the ratio of ionising photon number flux (the number of ionising photons impinging a unit area per unit time) to particle (atom + ion) number density; $q=\mathcal{U}\times c$, where $\mathcal{U}$ and $c$ are the dimensionless ionisation parameter and speed of light, respectively \citep{Dopita_2013_ApJS_208_10,Blanc_2015_ApJ_798_99,Mingozzi_2020_A&A_636_A42}.
For photoionisation model grids, IZI employs several model grids calculated using photoionisation codes such as CLOUDY \citep{Ferland_1998_PASP_110_761,Ferland_2017_RMxAA_53_385} or MAPPINGS \citep{Sutherland_1993_ApJS_88_253,Allen_2008_ApJS_178_20,Dopita_2013_ApJS_208_10}.
IZI has been updated further, such as by changing the posterior sampling method \citep{Congiu_2023_A&A_672_A148}.

% Bayes' theorem, likelihood definition
% why epsilon is needed
To obtain the estimates of \metal{}, \logq{} and \EBV{}, IZI performs Bayesian parameter estimation based on Bayes' theorem \citep{Sivia_2006_book,Toussaint_2011_RvMP_83_943,Sharma_2017_ARA&A_55_213}, given as
\begin{eqnarray} \label{eq:bayes}
    p(H|D,I) & = &\frac{p(D|H,I)\,p(H|I)}{p(D|I)}
\end{eqnarray}
or
\begin{eqnarray} \label{eq:name}
    \text{Posterior} & = & \frac{\text{Likelihood}\times\text{Prior}}{\text{Evidence}},
\end{eqnarray}
where $p(\cdot|\cdot)$, $H$, $D$, and $I$ denote the conditional probability, hypothesis (model parameters), data, and background information, respectively.
The evidence is a constant quantity independent of the model parameters; hence the likelihood and prior determine the shape of the posterior distribution along the model parameters.

The likelihood is calculated in IZI by comparing the emission line fluxes from the observation and photoionisation model grid, assuming a Gaussian distribution of each emission line flux \citep{Blanc_2015_ApJ_798_99,Mingozzi_2020_A&A_636_A42}.
The likelihood has the same form in both IZI versions of \cite{Blanc_2015_ApJ_798_99} and \cite{Mingozzi_2020_A&A_636_A42} and is given as
\begin{eqnarray} \label{eq:likelihood}
    p(D|H,I) = \prod_{i=1}^{n} \frac{1}{\sqrt{2\pi \left(e_i^2 + \epsilon^2\,f_i'^2 \right)}}\,\mathrm{exp} \left[ -\frac{\left(f_i - f_i'\right)^2}{2 \left(e_i^2 + \epsilon^2\,f_i'^2 \right)} \right],
\end{eqnarray}
where $f_i$, $e_i$, $f_i'$, $i$, and $n$ indicate the observed line flux, observed line flux error\footnote{Although `uncertainty' may be more accurate than the term `error,' we use `error' in this paper for consistency with earlier descriptions of IZI.}, reddened model line flux, index for each emission line, and total number of emission lines, respectively.
The parameter $\epsilon$ is a systematic fractional error term that represents the uncertainty of the emission line flux from the photoionisation model grid; this stems from the fact that the metallicities estimated using the photoionisation models differ from those estimated using empirical diagnostics \citep{Kewley_2008_ApJ_681_1183,Dopita_2013_ApJS_208_10,Blanc_2015_ApJ_798_99}.
Taking into account this metallicity discrepancy, \cite{Blanc_2015_ApJ_798_99} adopted 0.1 dex for $\epsilon$, and \cite{Mingozzi_2020_A&A_636_A42} modified $\epsilon$ for \Ha{} as 0.01 dex to give \Ha{} more weight than the other lines when estimating the model parameters.
When calculating the likelihood, IZI uses the normalised line fluxes from the observation and photoionisation model grid instead of the line fluxes, because it requires an additional scaling model parameter.
The observed emission line fluxes are normalised by the \Hb{} flux or the strongest line flux when \Hb{} is not available.
The reddened model emission line fluxes are correspondingly normalised.
IZI provides diverse photoionisation model grids, including the model calculations of \cite{Levesque_2010_AJ_139_712,Dopita_2013_ApJS_208_10,PerezMontero_2014_MNRAS_441_2663,Byler_2017_ApJ_840_44}.
These model grids provide normalised, intrinsic emission line flux for given \metal{} and \logq{}.
The reddened model emission line fluxes are calculated from these intrinsic line fluxes for a given \EBV{} using the Calzetti attenuation law \citep{Calzetti_2000_ApJ_533_682}.

By default, the priors are set as uniform in IZI.
The parameter ranges of \logq{} and \metal{} are limited to the parameter coverages of the photoionisation model grids, while that of \EBV{} is limited to (0, 1).
IZI also provides an option for the priors of \logq{} and \metal{} to have a Gaussian distribution.
\cite{Mingozzi_2020_A&A_636_A42} used this Gaussian prior to restrict \logq{} with the line ratio of \SoSfull{} from observations.
We discuss two issues related to this prior setting (the mismatch of line flux terms and violation of Bayes' theorem) in Section \ref{sec:prior}.

% MCMC
The posterior, which is proportional to (likelihood $\times$ prior), is determined by the Markov chain Monte Carlo (MCMC) sampling in IZI.
The MCMC method determines the shape of the posterior by randomly sampling the posterior, visiting regions with higher probability more frequently \citep{Hogg_2018_ApJS_236_11,Sharma_2017_ARA&A_55_213,Toussaint_2011_RvMP_83_943}.
IZI employs the affine-invariant ensemble sampler called emcee for MCMC sampling \citep{ForemanMackey_2013_PASP_125_306,ForemanMackey_2019_JOSS_4_1864}.
When the sampling is finished, IZI returns the resultant samples of the three model parameters---\metal{}, \logq{}, and \EBV{}---from which we can plot the posterior distribution.

\section{Parameter Recovery Tests of IZI and Results}

\begin{table*}
    \centering
    \caption{Mock data configuration and corresponding specifications. \label{tbl:config}}
    \begin{tabular}{lrccccl}
        \hline
        {name} & {SNR$^a$} & {log $\epsilon$$^b$} &  {\metal$^c$} & {log $q$$^d$} & {\EBV$^e$} & {modelled emission lines $^f$} \\
        \hline
        \input{tbl/mock_test_setup}\\
        \hline
        \multicolumn{7}{l}{\parbox{16cm}{$^a$ The signal-to-noise ratio of the emission lines defined as the ratio of emission line flux to its uncertainty.}} \\
        \multicolumn{7}{l}{\parbox{16cm}{$^b$ The logarithm of the systematic fractional error in the emission line flux predicted by the photoionisation model grid. This determines the amount of model error.}} \\
        \multicolumn{7}{l}{\parbox{16cm}{$^c$ The metallicity expressed in terms of the oxygen abundance \citep{Maiolino_2019_A&ARv_27_3}.}} \\
        \multicolumn{7}{l}{\parbox{16cm}{$^d$ The logarithm of the ionisation parameter, which is defined as the ratio of ionising photon number flux (the number of ionising photon impinging a unit area per unit time) to particle (atom + ion) number density. See text for more details.}} \\
        \multicolumn{7}{l}{\parbox{16cm}{$^e$ The colour excess for the emission lines.}} \\
        \multicolumn{7}{l}{\parbox{16cm}{$^f$ The wavelength tags are quoted as those given in IZI. The \Hb{} line is always included by default since all line fluxes are normalised by the \Hb{} flux in the photoionisation model grid. The auroral lines \citep{Peimbert_2017_PASP_129_82001,PerezMontero_2017_PASP_129_43001,Maiolino_2019_A&ARv_27_3} are underlined.}} \\
        \multicolumn{7}{l}{\parbox{16cm}{$^\dag$ The specification is the same with the corresponding `snr010' case, except for the SNR; for example, the specifications for `mock\_snr010\_a' and `mock\_snr020\_a' are the same except for the SNR. The corner plots for the `snr020' and 'snr003' cases are given as online supplementary figures.}} \\
        \multicolumn{7}{l}{\parbox{16cm}{$^\ddag$ The parameter \logep{} for \Ha{} was set as 0.01 following \cite{Mingozzi_2020_A&A_636_A42} (see the text). }} \\
    \end{tabular}
\end{table*}

%x model grid line flux given in ratio to the H-beta line
%x SNR means the SNR of line flux ratios for each emission line relative to H_beta
%x used grid: Dopita_2013_ApJS_208_10
To check the reliability of the parameter estimates of IZI, we ran IZI on several mock datasets with known parameter estimates.
Those mock datasets were generated by IZI under different model configurations.
Table \ref{tbl:config} shows the designed mock data configurations.
All parameter recovery tests used the photoionisation model grid of \cite{Dopita_2013_ApJS_208_10} with the Maxwell-Boltzmann electron energy distribution ($\kappa=\infty$), the model grid used in the MaNGA data analysis of \cite{Mingozzi_2020_A&A_636_A42}.
We varied the signal-to-noise ratio (SNR) value and the systematic fractional flux error of the model (\logep, Eq.~(\ref{eq:likelihood})) to examine how they affect the parameter recovery of IZI.
The SNR values in Table \ref{tbl:config} are the ratios of the emission line flux to its uncertainty for each line. 
We selected the values of the three model parameters as follows: [gas metallicity \metal, ionisation parameter \logq, nebular emission-line colour excess \EBV] = (8.8, 7.1, 0.1).
The values for \metal{} and \logq{} were adopted considering the estimates of \cite{Mingozzi_2020_A&A_636_A42} for their MaNGA data of local ($z<0.08$) star-forming galaxies.
The low \EBV{} of 0.1 was deliberately selected to check the possibility of \EBV{} overestimation.
The combinations of the modelled emission lines were also varied to represent different datasets.

\begin{table}
    \centering
    \caption{Adopted priors for the parameter recovery tests \label{tbl:priors}}
    \begin{tabular}{cl}
        \hline
        {Parameter} & {Prior distribution$^\dag$}      \\
        \hline
        \metal{}    &   Uniform; (7.39, 9.39)$^a$    \\
        \logq{}     &   Uniform; (6.50, 8.50)$^a$    \\
        \EBV{}      &   Uniform; (0.0, 1.0)  \\
        \hline
        \multicolumn{2}{l}{\parbox{4.1cm}{$^\dag$ We followed the priors \cite{Mingozzi_2020_A&A_636_A42} used when applying IZI to their MaNGA data with a uniform \logq{} prior.}} \\
        \multicolumn{2}{l}{\parbox{4.1cm}{$^a$ This corresponds to the full range of the adopted photoionisation model grid of \cite{Dopita_2013_ApJS_208_10}.}} \\
    \end{tabular}
\end{table}

% how to sample posterior
With the mock data ready, we performed a Bayesian parameter estimation on the mock data using IZI.
The prior was set to be uniform on the allowed parameter range (Table \ref{tbl:priors}).
For more flexible analysis, we wrote our own scripts using functions and information provided by IZI.
This script was written as a Jupyter notebook and was shared online including instructions (see the section `Data Availability').
We confirmed that our script and IZI returned consistent results for the same mock data.
Our script additionally includes a procedure for monitoring the MCMC sampling convergence, which enables us to assess how close the obtained samples are to the target posterior.
We considered the MCMC sampling to have converged enough when the sampling generated a suitable number of independent samples.
The number of independent samples was estimated by the quantity called effective sample size (ESS), and we stopped the MCMC sampling when all parameters achieved ESSs of $>$ 2,000.
The quantity ESS is defined as N/\tauint{}, where N and \tauint{} represent the sample length and integrated autocorrelation time, respectively \citep{Sharma_2017_ARA&A_55_213}. The integrated autocorrelation time (\tauint) is expressed as follows:
\begin{equation} \label{eq:iat}
    \tau_{\text{int}}=\sum^{\infty}_{t=-\infty}\rho_{xx}(t)
    \text{, where } \rho_{xx}(t)=\frac{\mathbb{E}[(x_i-\bar{x})(x_{i+t}-\bar{x})]}{\mathbb{E}[(x_i-\bar{x})^2]},
\end{equation}
where $\rho_{xx}$ represents the autocorrelation function for the sequence $\{x_i\}$, $t$ denotes the time difference---or distance---between two points in the sequence $\{x_i\}$, $\bar{x}$ represents the mean of the sequence $\{x_i\}$, and $\mathbb{E}\left[\cdot\right]$ denotes the expectation value.
We excluded the samples in the `burn-in' phase when monitoring the convergence and performing the parameter estimation.
The readers may refer to \cite{Shinn_2020_MNRAS_499_1073} and \cite{Shinn_2022_MNRAS_517_474} for more details on our MCMC sampling.

\begin{figure*}
	\includegraphics[scale=0.5]{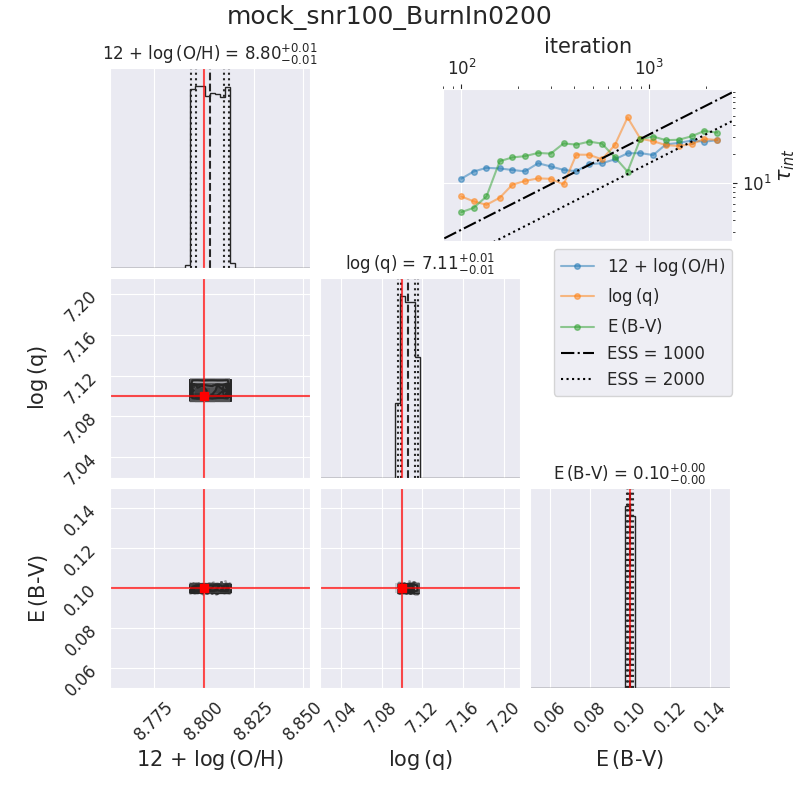}
    \caption{Posterior distribution and evolution of the integrated autocorrelation times (\tauint) for the mock data `mock\_snr100' (see Table~\ref{tbl:config}). Six panels in the lower-left corner show the corner plots of the Markov chain Monte Carlo (MCMC) sampling results, demonstrating the correlations among the model parameters and their marginal distributions. Solid red lines represent input values used to generate the mock data. One vertical dashed and four vertical dotted lines indicate the median, 1-$\sigma$ (68 \%) and 2-$\sigma$ (95 \%) credible intervals, respectively. The panel in the upper-right corner shows the evolution of \tauint. We plot two straight lines (dot-dashed and dotted) for convergence diagnosis, corresponding to effective sample sizes (ESSs) of 1000 and 2000, respectively. The title at the top of the figure indicates the following---snr\#\#\#: signal-to-noise ratio of the mock data and BurnIn\#\#\#\#: burn-in iterations excluded before plotting.}
    \label{fig:mock_snr100}
\end{figure*}

% epsilon of set_dex = 10^set_dex
% 1. SNR = 100, no model error
First, we created mock data to inspect how IZI pinpoints the three input parameter values of \metal{}, \logq{}, and \EBV{}.
We excluded the model error (\logep{} $=-\infty$, i.e., $\epsilon=0$), set a significantly high SNR of 100, and included all emission lines in the model grid; we selected this SNR of 100 considering the typical SNR's higher end for the emission lines of local star-forming galaxies observed in the MaNGA survey \citep{Mingozzi_2020_A&A_636_A42}.
This mock data are denoted as \mhund{} (see Table \ref{tbl:config}), and Fig.~\ref{fig:mock_snr100} shows the result of the parameter recovery test.
As shown in Fig.~\ref{fig:mock_snr100}, IZI recovers the input parameter values used to generate the corresponding mock data well.
The correlation panels (off-diagonal panels of the corner plot) show rectangular distributions.
This is because the posterior is dominated by one model grid cell as designed, due to the minimal uncertainty term in the likelihood (Eq.~(\ref{eq:likelihood})).

\begin{figure*}
	\includegraphics[scale=0.5]{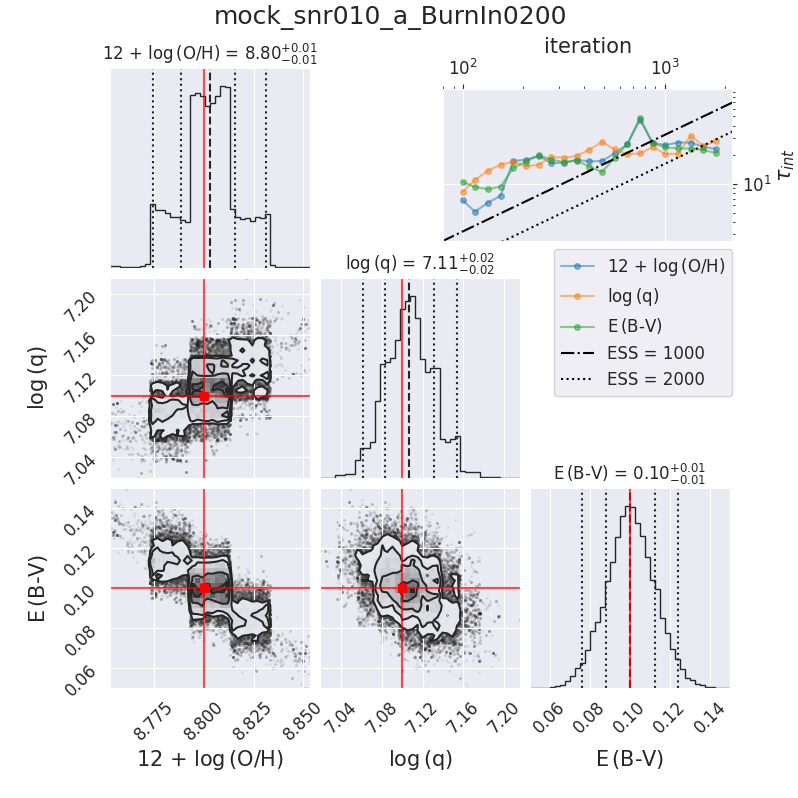}
    \caption{Posterior distribution and evolution of the integrated autocorrelation times (\tauint) for the mock data `mock\_snr010\_a' (see Table~\ref{tbl:config}). The figure description is otherwise the same as that for Fig.~\ref{fig:mock_snr100}. The plots for the `snr020' and `snr003' cases are given as online supplementary figures.}
    \label{fig:mock_snr010_a}
\end{figure*}

% 2. SNR = 10, no model error
Second, we created the same mock data as \mhund{} except for a smaller SNR of 10 instead of 100, which is called \mtena{} (see Table \ref{tbl:config}).
We selected this SNR of 10 considering the typical SNR's lower end for the emission lines of local star-forming galaxies observed in the MaNGA survey \citep{Mingozzi_2020_A&A_636_A42}.
Fig.~\ref{fig:mock_snr010_a} shows the result of the parameter recovery test.
As shown in Fig.~\ref{fig:mock_snr010_a}, IZI recovers the input parameter values within the 1-$\sigma$ credible interval well.
The distributions seen in the correlation panels have a more diffuse shape than those of \mhund{}, which is natural because the mock data \mtena{} has a lower SNR.
They also show that the three model parameters are not independent: the parameters \metal{} and \logq{} show a positive correlation, while the other two pairs show negative correlations.

\begin{figure*}
	\includegraphics[scale=0.5]{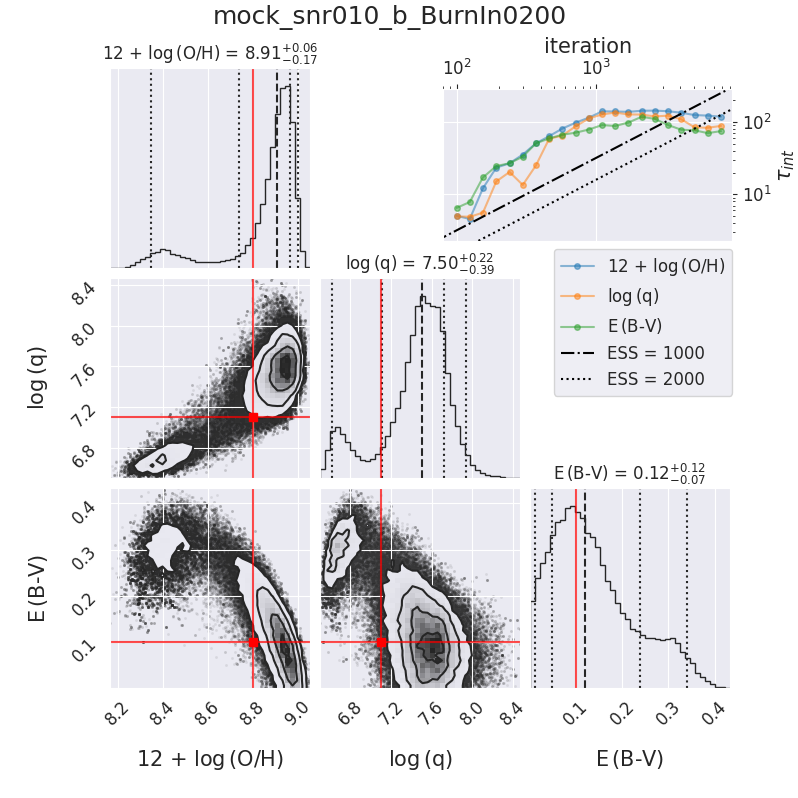}
    \caption{Posterior distribution and evolution of the integrated autocorrelation times (\tauint) for the mock data `mock\_snr010\_b' (see Table~\ref{tbl:config}). The figure description is otherwise the same as that for Fig.~\ref{fig:mock_snr100}. The plots for the `snr020' and `snr003' cases are given as online supplementary figures.}
    \label{fig:mock_snr010_b}
\end{figure*}

% epsilon of set_dex = 10^set_dex
% 3. SNR = 10, with model error
Third, we created the same mock data as \mtena{} except for including the model error to examine its impact on the parameter recovery of IZI.
As in \cite{Mingozzi_2020_A&A_636_A42}, we included a model error of 0.1 dex (\logep{} = 0.1, i.e., $\epsilon=10^{0.1}$) for all emission lines except for \Ha{}, for which a model error of 0.01 dex was applied.
This mock data is called \mtenb{} (see Table \ref{tbl:config}), and Fig.~\ref{fig:mock_snr010_b} shows the result of the parameter recovery test.
The marginalised one-dimensional probability density functions (PDFs) show that \logq{} is overestimated by approximately 1-$\sigma$, while \metal{} and \EBV{} recover the input values within 1-$\sigma$.
Each correlation panel shows a distribution with two separate peaks.
These results indicate that the posterior is severely deformed by the inclusion of model error compared to the \mtena{} case.

\begin{figure*}
	\includegraphics[scale=0.5]{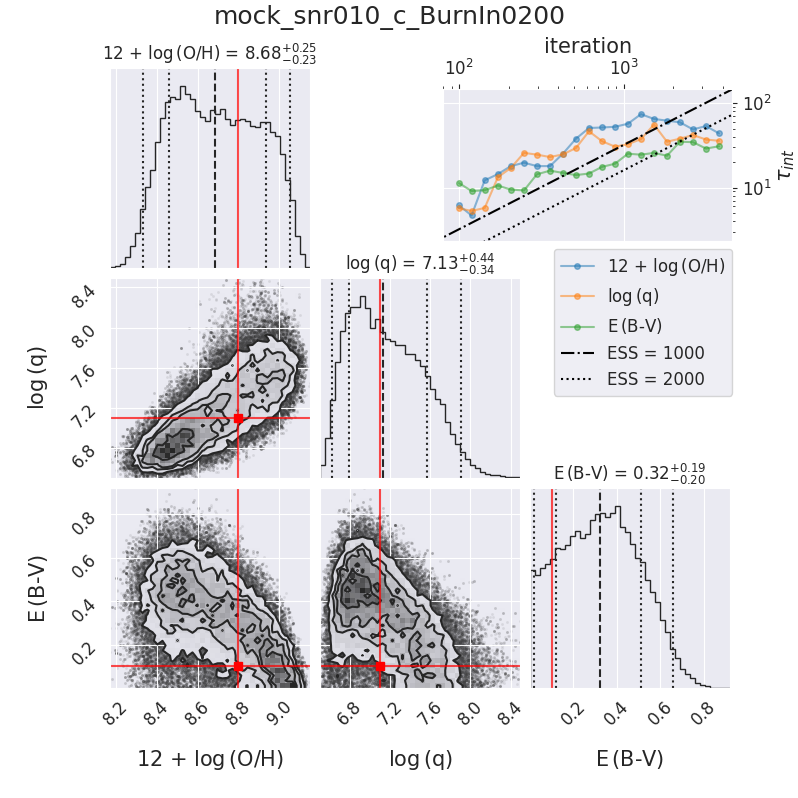}
    \caption{Posterior distribution and evolution of the integrated autocorrelation times (\tauint) for the mock data `mock\_snr010\_c' (see Table~\ref{tbl:config}). The figure description is otherwise the same as that for Fig.~\ref{fig:mock_snr100}. The plots for the `snr020' and `snr003' cases are given as online supplementary figures.}
    \label{fig:mock_snr010_c}
\end{figure*}

% 4. Minggozzi et al. setup
Fourth, we created the same mock data as \mtenb{} except for the number of modelled emission lines.
We called this mock data \mtenc{} and narrowed down the modelled emission lines as seen in Table \ref{tbl:config}.
This mock data configuration is the same as that used by \cite{Mingozzi_2020_A&A_636_A42} to analyse MaNGA data of local star-forming galaxies with uniform priors.
Fig.~\ref{fig:mock_snr010_c} shows the results of the parameter recovery test.
The marginalised PDFs indicate that \EBV{} is overestimated by approximately 1-$\sigma$, while \metal{} and \logq{} recover the input values within 1-$\sigma$.
Each correlation panel does not show two separate peaks as prominently as can be seen in the \mtenb{} results (Fig.~\ref{fig:mock_snr010_b}).
This seems to be due to the smaller number of emission lines used for parameter estimation, which gives looser constraints on the parameters.
The positive and negative correlations between \metal{}, \logq{}, and \EBV{} are the same as those in the \mtena{} results (Fig.~\ref{fig:mock_snr010_a}).

\begin{figure*}
	\includegraphics[scale=0.5]{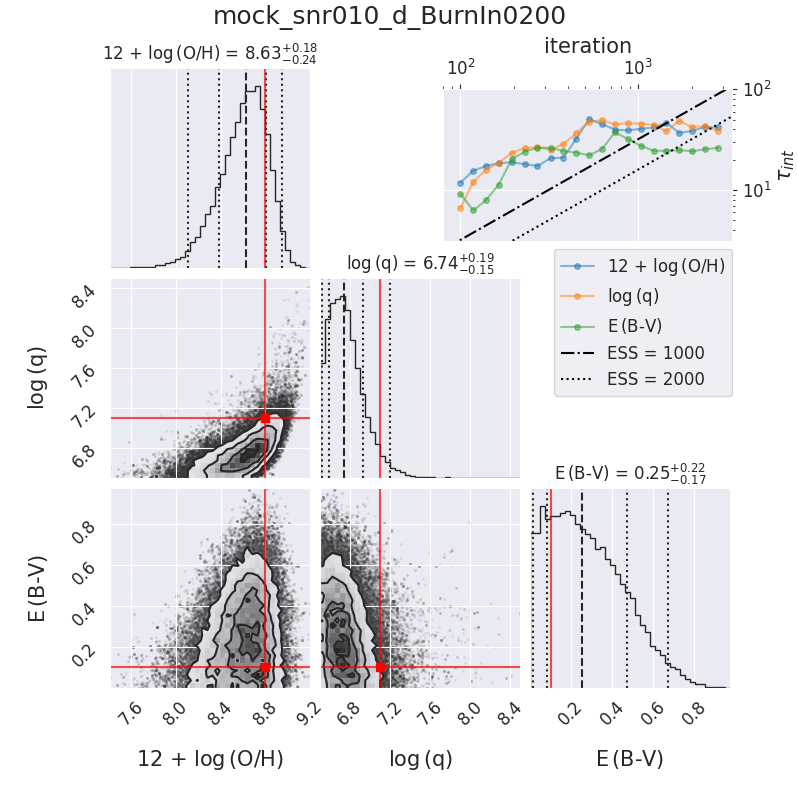}
    \caption{Posterior distribution and evolution of the integrated autocorrelation times (\tauint) for the mock data `mock\_snr010\_d' (see Table~\ref{tbl:config}). The figure description is otherwise the same as that for Fig.~\ref{fig:mock_snr100}. The plots for the `snr020' and `snr003' cases are given as online supplementary figures.}
    \label{fig:mock_snr010_d}
\end{figure*}

% 5. high-z setup, z <~ 0.6
Fifth, we created another variation of \mtenb{} by selecting a different set of emission lines that could be observed in optical for distant galaxies of redshift $z\loa0.6$, where the \Ha{} emission line falls outside the typical optical wavelength coverage, for example $\sim4,000-8,000$ \AA, but the [\ion{O}{III}] $\lambda$5007 line does not.
We called this mock data \mtend{} and selected the modelled emission lines as seen in Table \ref{tbl:config}.
The input metallicity of \metal{} = 8.8 for the mock data does not seem to be unrealistic, considering the metallicity of star-forming galaxies at $z\sim0.6$ \citep{Lilly_2003_ApJ_597_730,Kobulnicky_2004_ApJ_617_240}.
Fig.~\ref{fig:mock_snr010_d} shows the results of the parameter recovery test.
The marginalised PDFs show that \logq{} is underestimated by more than 1-$\sigma$, while the input values for \metal{} and \EBV{} are recovered within 1-$\sigma$.
The pair \metal{} and \logq{} shows a positive correlation, while the other two pairs show almost no correlation.

\begin{figure*}
	\includegraphics[scale=0.5]{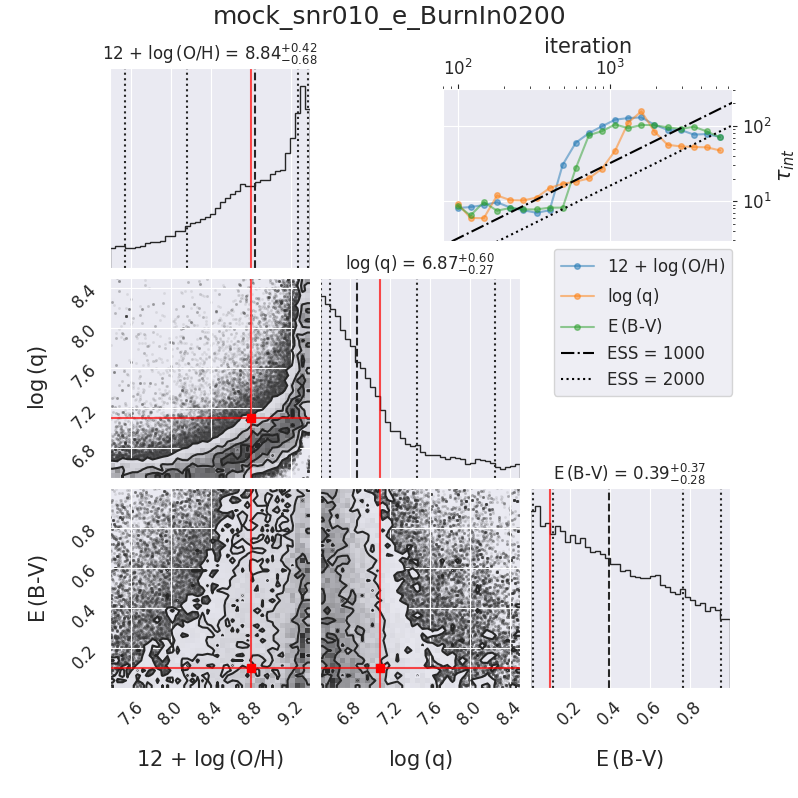}
    \caption{Posterior distribution and evolution of the integrated autocorrelation times (\tauint) for the mock data `mock\_snr010\_e' (see Table~\ref{tbl:config}). The figure description is otherwise the same as that for Fig.~\ref{fig:mock_snr100}. The plots for the `snr020' and `snr003' cases are given as online supplementary figures.}
    \label{fig:mock_snr010_e}
\end{figure*}

% 6. Oxygen line only
Lastly, we created another variation of \mtenb{} by selecting the emission lines used for the strong line diagnostics R23 $\equiv(\text{[\ion{O}{II}]}\,\lambda3726 + \text{[\ion{O}{II}]}\,\lambda3729 + \text{[\ion{O}{III}]}\,\lambda4959 + \text{[\ion{O}{III}]}\,\lambda5007)/\text{H}\,\beta$ (\citealt{Maiolino_2019_A&ARv_27_3}).
We called this mock data \mtene{}, and Fig.~\ref{fig:mock_snr010_e} shows the results of the parameter recovery test.
The marginalised PDFs indicate that \EBV{} is overestimated by approximately 1-$\sigma$, while \metal{} and \logq{} recover the input values within 1-$\sigma$.
The PDFs also have monotonic distributions and wider 1-$\sigma$ credible intervals than those of previous mock datasets.
The correlation panels show diffuse two-dimensional distributions and no signs of strong positive or negative correlations.
These posterior features, distinctive to other mock-data cases, seem to be caused by the smallest number of emission lines used for the analysis.

\begin{figure*}
	\includegraphics[scale=0.5]{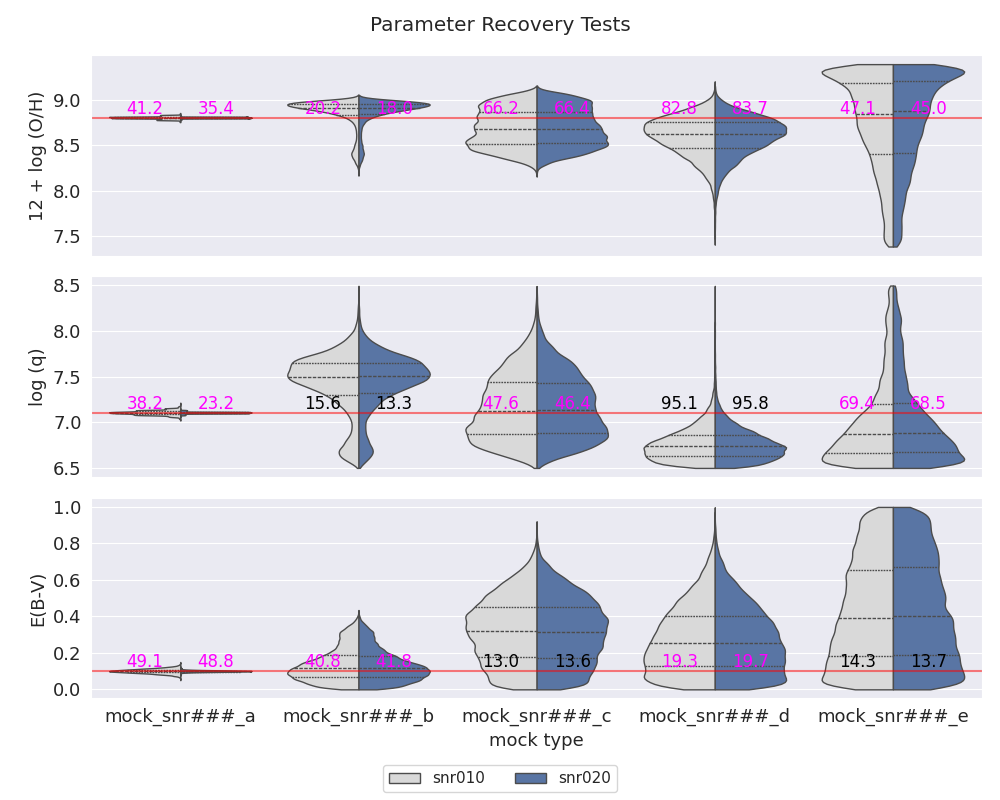}
    \caption{Comparison between the parameter posterior distributions of the SNR=10 and SNR=20 cases. The abscissas represent the mock data configuration name (see Table~\ref{tbl:config}), and the ordinates represent the model parameters. The shadings differ for different signal-to-noise ratios (SNRs) used to create the mock data (see Table~\ref{tbl:config}). Horizontal black dashed and dotted lines indicate the median and quartile of the distribution, respectively. The horizontal red lines indicate the input values of the parameters used to create the mock data. The cumulative percentage at the input values, that is, $P(y\le y_{\textrm{in}})$, is denoted on the horizontal red lines to show how well the posterior recovers the input values. The number is indicated in magenta when the input value falls within the 1-$\sigma$ credible interval, that is, $15.865<P(y\le y_{\textrm{in}})<84.135$.}
    \label{fig:violinplots020}
\end{figure*}

\begin{figure*}
	\includegraphics[scale=0.5]{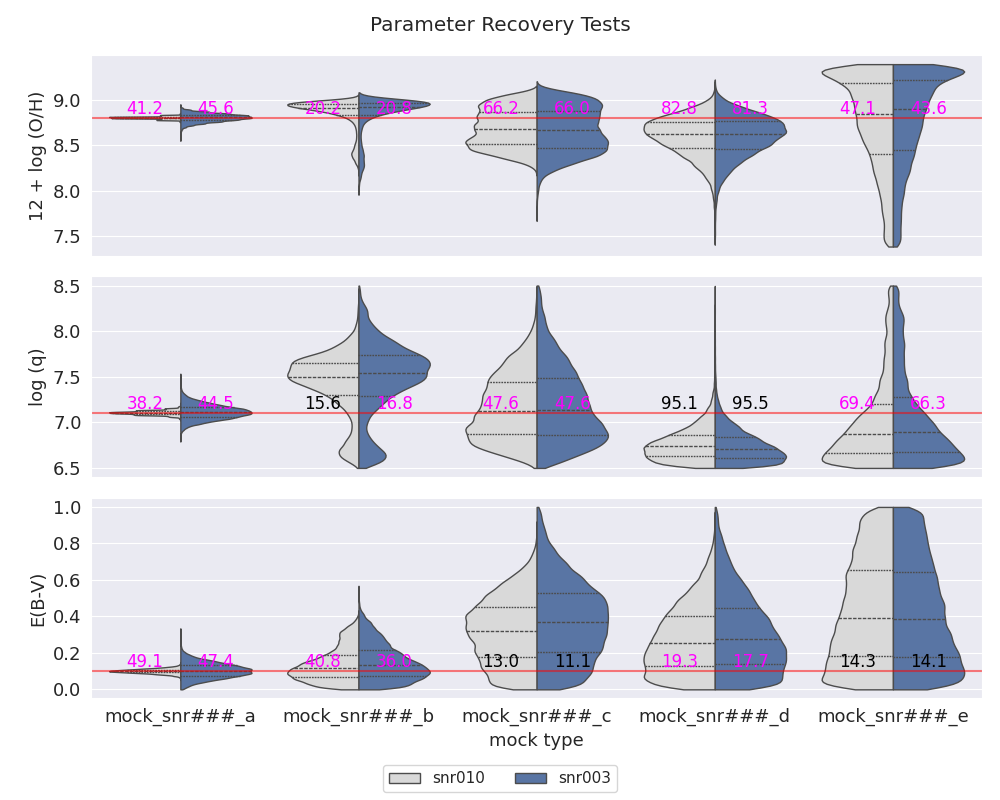}
    \caption{Comparison between the parameter posterior distributions of the SNR=10 and SNR=3 cases. The figure description is otherwise the same as that for Fig.~\ref{fig:violinplots020}.}
    \label{fig:violinplots003}
\end{figure*}

% 7. SNR=20 & SNR=3 case
Additionally, we extended the above analyses by changing the SNR of the mock data from 10 to 20 and 3 to check how the SNR affects the parameter recovery of IZI.
Figs.~\ref{fig:violinplots020} and \ref{fig:violinplots003} summarise the comparison of the SNR = 10 case to the SNR = 20 and SNR = 3 cases, respectively\footnote{The corner plots for the SNR = 20 and SNR = 3 cases are available in the online supplementary data (see the `Supporting Information' section).}.
As the symmetric (left-right) shape of the marginalised PDFs shown in Fig.~\ref{fig:violinplots020} indicates, the SNR = 10 and SNR = 20 cases recover the input values in a similar manner when the model error is included (\logep{} = 0.1), that is, for the four mock data `mock\_snr\#\#\#\_[b-e]'.
By contrast, the two cases show different marginalised PDFs when the model error is excluded (\logep{} = -$\infty$, i.e., $\epsilon=0$), that is, for the mock data `mock\_snr\#\#\#\_a'.
These facts mean that the model error dominates the data error for both the SNR = 10 and SNR = 20 cases if the model error is included.
Fig.~\ref{fig:violinplots020} also demonstrates that IZI does not recover one input value within the 1-$\sigma$ credible interval when the model error is included, that is, for `mock\_snr\#\#\#\_[b-e]'.
The comparison plot of the SNR = 10 and SNR =3 cases (Fig.~\ref{fig:violinplots003}) has approximately the same characteristics as those of the SNR = 10 and SNR =20 cases (Fig.~\ref{fig:violinplots020}).
When the model error is excluded (`mock\_snr\#\#\#\_a'), the PDFs of the SNR = 3 case show a wider width than that of the SNR = 10 case, which is due to its lower SNR.
When the model error is included (`mock\_snr\#\#\#\_[b-e]'), the PDFs of the SNR = 3 cases become slightly wider than that of the SNR = 10 case, but the symmetric (left-right) shape of the PDFs still remains in Fig.~\ref{fig:violinplots003}; due to this broadening, \logq{} recovers the input parameter values within the 1-$\sigma$ credible interval in the `mock\_snr003\_b' case, being different from the `mock\_snr010\_b' and `mock\_snr020\_b' cases.
The symmetry seen in Fig.~\ref{fig:violinplots003} indicates that the model error is still dominant in determining the width of PDFs even when the SNR is as low as 3.

\section{Discussion}
\subsection{Parameter Recovery Test Results}
% general message: be careful when including model error in the likelihood
% extension: same thing might happen to any likelihoods that includes the model error
% bad retracing may not affect the gradient of some physical quantities

The corner plots of the three mock data (\mhund, \mtena{} and \mtenb{}; Figs.~\ref{fig:mock_snr100}-\ref{fig:mock_snr010_b}) clearly show how the model error changes the posterior's shape and hinders correct parameter estimation.
When the model error is not included in the likelihood (see Figs.~\ref{fig:mock_snr100} and \ref{fig:mock_snr010_a}), the posterior recovers the input parameter values well; it has the highest probability around the input values of \metal{}, \logq{}, and \EBV{} and spreads slightly as the SNR decreases.
By contrast, the posterior is transformed to a completely different shape when the model error is included in the likelihood (see Figs.~\ref{fig:mock_snr010_a} and \ref{fig:mock_snr010_b}).
As a result, the ionisation parameter (\logq{}) is overestimated by up to 1-$\sigma$ ($\sim$ 0.4 dex).
This overestimation is more interesting when considering that the three mock datasets use all emission lines provided in the model grid of \cite{Dopita_2013_ApJS_208_10} (Table \ref{tbl:config}).
This means that overestimation is inevitable even when all possible emission line information is used.

The mock data \mtenc{} uses a smaller number of emission lines than that of \mtenb{} (Table \ref{tbl:config}), but the line selection is the same as that used by \cite{Mingozzi_2020_A&A_636_A42} to analyse their MaNGA data of local star-forming galaxies.
In this sense, we can regard \mtenc{} as a more realistic mock data.
As shown in Fig.~\ref{fig:mock_snr010_c}, the nebular emission-line colour excess (\EBV{}) is overestimated in this case.
The overestimation is as large as 1-$\sigma$ ($\sim$ 0.2 mag).
If we use this overestimated \EBV{} for the line-flux correction, \Ha{} and [\ion{O}{II}] $\lambda$3727 (= [\ion{O}{II}] $\lambda$3726 + [\ion{O}{II}] $\lambda$3729) line fluxes are overestimated by a factor of $\sim$ 2 and $\sim$ 3, respectively; consequently, this can lead to the star-formation rate (SFR) overestimation by the same factors, since the SFR estimate linearly scales with the line luminosity \citep{Kennicutt_1998_ARA&A_36_189,Kennicutt_2012_ARA&A_50_531}.
\cite{Mingozzi_2020_A&A_636_A42} presented their MaNGA data results analysed under the same line selection as that for \mtenc{} and the same prior choice as shown in Table~\ref{tbl:priors}, although their presentation was for demonstration purposes\footnote{\cite{Mingozzi_2020_A&A_636_A42} focused on the results of their MaNGA data analysis with the Gaussian prior on the ionisation parameter (\logq{}). However, as discussed in Section \ref{sec:prior}, we found that this Gaussian prior setup needs to be handled carefully.}.
Taking into account our \mtenc{} test results, their estimate for \EBV{} could be biased.

The mock data \mtend{} represents a spectrum of distant galaxies at $z\loa0.6$ that falls in optical wavelengths (Table \ref{tbl:config}).
Its corner plot (Fig.~\ref{fig:mock_snr010_d}) shows that the ionisation parameter (\logq{}) is underestimated by more than 1-$\sigma$ ($\sim$ 0.4 dex).
The metallicity (\metal{}) and the colour excess (\EBV{}) are estimated within the 1-$\sigma$ credible interval, but they are marginal.
These suggest that the IZI analysis results for distant ($z\loa0.6$) galaxies could be biased.

The mock data \mtene{} represents a case where only a few emission lines are detected (Table \ref{tbl:config}), and we selected the emission lines used in the strong line diagnostics R23 (\citealt{Maiolino_2019_A&ARv_27_3}).
The corner plot (Fig.~\ref{fig:mock_snr010_e}) shows the most unconstrained one-dimensional PDFs---none have a single-peaked shape; the mode of each PDF is located at either end of the plot range; and the 1-$\sigma$ credible intervals are the widest among all the results of the parameter recovery test.
These properties seem to be due to the few emission lines used in the mock data.
In particular, the colour excess (\EBV{}) is overestimated by up to 1-$\sigma$ ($\sim$ 0.3 mag) despite the wide width of its PDF.
This \EBV{} overestimate of 0.3 mag can lead to SFR overestimation by factors of 3 and 5 when using \Ha{} and [\ion{O}{II}] $\lambda$3727 emission lines, respectively \citep{Kennicutt_1998_ARA&A_36_189,Kennicutt_2012_ARA&A_50_531}.
These results indicate that IZI estimates from small number ($\sim5$ or so) of emission lines could be significantly biased.

Figs.~\ref{fig:violinplots020} and \ref{fig:violinplots003} present the posterior comparison of the SNR = 10 case to the SNR = 20 and SNR =3 cases, respectively.
When the model error is included in the likelihood term (`mock\_snr\#\#\#\_[b-e]'), one model parameter does not recover the input parameter value within the 1-$\sigma$ credible interval, except for the `mock\_snr003\_b' case.
This indicates that we cannot rule out the possibility of biased parameter estimation of IZI when the model error is included.
The PDFs are asymmetric (left-right) when the model error is not included (`mock\_snr\#\#\#\_a'), while they are almost symmetric when the model error is included (`mock\_snr\#\#\#\_[b-e]', see Table \ref{tbl:config}).
The symmetry becomes slightly weaker in Fig.~\ref{fig:violinplots003}, because the increased data error slightly widens the PDF width in the SNR = 3 case; however, the overall shape of the PDFs is similar to the SNR = 10 case.
This symmetry means that the model error is significantly larger than the data error and is a dominant factor in determining the posterior's shapes.

The above underestimation and overestimation examples are based on the parameter recovery test configurations we adopted (Table \ref{tbl:config}), and different estimation results will be obtained when using different configurations.
Therefore, we should regard our parameter recovery test results as a guide for estimating the parameters with IZI.
We also strongly recommend that IZI users run parameter recovery tests in the configuration suitable for their data to check the reliability of the IZI estimates.
To facilitate the running of such parameter recovery tests, we have shared the Jupyter script used for our parameter recovery tests online with instructions (see the `Data Availability' section).

We note three points related to our parameter recovery tests.
First, the underestimation or overestimation might not affect some physical quantities in some cases.
For instance, the radial metallicity gradient will not change if the metallicity is underestimated or overestimated by the same amount along the radial direction.
Second, we found that NebulaBayes, a Bayesian analysis tool for the emission lines from photoionisation \citep{Thomas_2018_ApJ_856_89}, includes the same type of model error in the likelihood term as in IZI. 
Therefore, we recommend that users of NebulaBayes run parameter recovery tests before interpreting the NebulaBayes estimates, especially when studying the relationship between the model parameters \citep{Garner_2025_ApJ_978_70}, since even weakly biased estimates could result in a significant correlation between the model parameters.
Third, the causes of the underestimation and overestimation presented here are the meeting of the following three factors: (i) the inclusion of model error in the likelihood term, (ii) substantial model error compared to the data error, and (iii) variation of model error along with the model parameters.
This means that any parameter estimation satisfying these three conditions can underestimate or overestimate the parameters.
Therefore, parameter recovery tests are beneficial, especially when model error is considered in parameter estimation.

\subsubsection{Auroral Lines, Original IZI, and Photoionisation Model}
The auroral lines, the emission lines coming from high quantum levels, can play a crucial role in breaking degeneracies between model parameters since their excitation is sensitive to temperature \citep{Peimbert_2017_PASP_129_82001,Maiolino_2019_A&ARv_27_3}.
Therefore, the biased parameter estimation could be caused by the absence of the auroral lines rather than the model error.
We checked whether this is the case for the mock dataset `mock\_snr010\_c', which does not include any auroral lines (see Table \ref{tbl:config}).
We made another mock data configuration `mock\_snr010\_c\_ep0', which is the same as `mock\_snr010\_c', except for \logep{} = $-\infty$ (no model error).
IZI found the input values near the medians of its marginalised PDFs.\footnote{We provide the corner plot for `mock\_snr010\_c\_ep0' as an online supplementary data.}
This means that the biased parameter estimation seen for `mock\_snr010\_c' is not due to the absence of auroral lines but to the model error.
However, the absence of auroral lines could still be one of the causes of biased parameter estimation under other mock data configurations.

As mentioned in Section \ref{sec:izi}, the original IZI of \cite{Blanc_2015_ApJ_798_99} is different from the modified IZI of \cite{Mingozzi_2020_A&A_636_A42} in two aspects: (1) the original IZI does not have the parameter \EBV{} and (2) assumes the model error term $\epsilon$ of 0.1 dex for all emission lines.
This difference could cause the original IZI to suffer from less or no biased parameter estimation.
To test this possibility, we made other mock data configurations `mock\_snr010\_[b-e]\_Blanc', which are the same as 'mock\_snr010\_[b-e]', except for the two aspects mentioned above, respectively.
To mimic the absence of the parameter \EBV{}, we set the \EBV{} prior to be uniform over a very narrow range around the input value, that is (0.099, 0.101).
The results show that the parameters \metal{} and \logq{} are not always recovered within 1-$\sigma$.\footnote{We provide the corner plots for `mock\_snr010\_[b-e]\_Blanc' as an online supplementary data.}
These results suggest that the original IZI of \cite{Blanc_2015_ApJ_798_99} is probably also not free from biased parameter estimation.

% bayesian modeling, density structure
IZI returns its parameter estimates through the Bayesian parameter estimation framework.
Basically, it compares the observed emission line ratios to the corresponding ratios from the precalculated photoionisation model grids in the likelihood term (Eq.~(\ref{eq:likelihood})), and hence, its estimation accuracy depends on the accuracy of the photoionisation model.
The inconsistency between the gas metallicity estimates from the photoionisation model method and other methods (the direct electron temperature method and metal recombination line method) is well known \citep{Kewley_2008_ApJ_681_1183,Dopita_2013_ApJS_208_10,Blanc_2015_ApJ_798_99,Peimbert_2017_PASP_129_82001,Maiolino_2019_A&ARv_27_3}, and the inaccuracy of photoionisation models stems from various sources \citep{Peimbert_2017_PASP_129_82001,Maiolino_2019_A&ARv_27_3}; for example, the structural complexity of photoionised regions, such as density variations \citep[e.g.,][]{MendezDelgado_2023_MNRAS_523_2952}, deposition of mechanical energy, cosmic rays, assumed abundance relation between N and O, and dust depletion.
\deleted{The estimation accuracy of the photoionisation model is believed to be lower than that of the direct electron temperature method when considering the consistency with the metallicities of young stars (\citeauthor{Maiolino_2019_A&ARv_27_3} \citeyear{Maiolino_2019_A&ARv_27_3}).}
The amount of this inaccuracy is implemented with the parameter $\epsilon$ in IZI (Eq.~(\ref{eq:likelihood})).
If the accuracy of photoionisation models is improved, we can adopt a smaller $\epsilon$ and therefore obtain more accurate estimates with IZI.

\subsection{Prior for Ionisation Parameter Based on the \SoS{} ratio} \label{sec:prior}
% error propagation when normalizing the flux ==> No need to mention?
% empirical Bayes method, caution when using observation data in setting prior
For their MaNGA data analysis, \cite{Mingozzi_2020_A&A_636_A42} set a Gaussian prior on the ionisation parameter (\logq) using the observed line fluxes and calibration equation obtained from the photoionisation model of \cite{Diaz_1991_MNRAS_253_245}, which is
\begin{equation} \label{eq:prior_q}
    \mathrm{log}\,q = -1.68 \times \mathrm{log}(\SoSinv) + 7.49,
\end{equation}
where \SoSinv{} is the flux ratio of the corresponding emission lines, [\ion{S}{II}] $\lambda$6717, $\lambda$6731 and [\ion{S}{III}] $\lambda\lambda$9068, 9532.
However, we believe that this prior setting needs to be reconsidered for two separate reasons.

First, there is a mismatch in the line flux terms that appears in Eq.~(\ref{eq:prior_q}).
Eq.~(\ref{eq:prior_q}) was derived from the photoionisation model of \cite{Diaz_1991_MNRAS_253_245}, which does not include any treatment of dust attenuation.
Therefore, the line flux ratio \SoS{} in Eq.~(\ref{eq:prior_q}) should be the \emph{dereddened, intrinsic} line fluxes; however, the \emph{observed} line fluxes were used in \cite{Mingozzi_2020_A&A_636_A42}.
If we set the \logq{} prior properly using Eq.~(\ref{eq:prior_q}), the terms in the equation should be as follows, 
\begin{equation} \label{eq:prior_q_mod}
    \mathrm{log}\,q = -1.68\,\times\,\mathrm{log} \left[ 
    \frac{k_{\mathrm{cor}}\left(\lambda_{[\ion{S}{II}]},\, E(B-V)\right) \times f_{\mathrm{obs}}\left([\ion{S}{II}]\right)}{k_{\mathrm{cor}}\left(\lambda_{[\ion{S}{III}]},\, E(B-V)\right) \times f_{\mathrm{obs}}\left([\ion{S}{III}]\right)}
    \right]\,+\,7.49,
\end{equation}
where $k_{\mathrm{cor}}$ is the dereddening factor, which depends on the wavelength of the emission line and colour excess \EBV{}, and $f_{\mathrm{obs}}$ is the observed emission line flux.
In this prior setting, the \logq{} prior becomes dependent on another model parameter \EBV{}.
This dependence could result in parameter estimates different from those reported by \cite{Mingozzi_2020_A&A_636_A42}.

Second, the observed emission-line fluxes were used to set the \logq{} prior\footnote{A similar type of prior setting was used in the study by \cite{Shinn_2022_MNRAS_517_474}. We also note here that there exists a method which uses the data $D$ for a prior setting. It is called `empirical Bayes' \citep{Gelman_2013_book} and can be viewed as an approximation to the hierarchical Bayesian analysis.}, but this treatment violates Bayes' theorem.
In Bayes' theorem, the data $D$ plays a role that updates the prior and leads to the posterior (see Eq.~(\ref{eq:bayes})).
Therefore, a circular logic occurs if the prior depends on the data $D$, that is, the prior depends on itself: prior = $f_0(D) = f_0(f_1(\mathrm{posterior, prior}))$, where $f_0$ and $f_1$ are certain functions.
However, this circular logic can be avoided by not using the same emission line in both the likelihood and prior, since many other emission line fluxes are available in \citeauthor{Mingozzi_2020_A&A_636_A42}'s analyses.
\cite{Mingozzi_2020_A&A_636_A42} ended up using $f_{\mathrm{obs}}\left([\ion{S}{II}]\right)$ in both the likelihood and prior by setting the \logq{} prior using $f_{\mathrm{obs}}\left([\ion{S}{II}]\right)$.
Therefore, if $f_{\mathrm{obs}}\left([\ion{S}{II}]\right)$ is excluded from the likelihood when setting the \logq{} prior with Eq.~(\ref{eq:prior_q_mod}), one can avoid violating Bayes' theorem.

We identify two issues relevant to the \logq{} prior setting (the mismatch of line flux terms and violation of Bayes' theorem) and propose how to avoid them.
We believe that IZI users will obtain more accurate estimates if they follow our remedies when using the observed emission line fluxes of [\ion{S}{II}] $\lambda$6717, $\lambda$6731 and [\ion{S}{III}] $\lambda\lambda$9068, 9532 for the \logq{} prior setting.

% \subsection{Comments on the Previous IZI-related Studies}
% IZI has been used in numerous studies and we leave comments on some of the IZI-related studies in terms of the possibility that the estimates with IZI might be under- or overestimated.

% \cite{Zhang_2017_MNRAS_466_3217} showed gas metallicity distributions as a function of the galactic radius.
% Although they used the original version of IZI \citep{Blanc_2015_ApJ_798_99}, we think their

% the impact of diffuse ionised gas on the emission-line ratios from star-forming galaxies \citep{Zhang_2017_MNRAS_466_3217}, 
% the evolution of the dust-to-metal ratios in galaxies \citep{DeVis_2019_A&A_623_A5},
% the metallicity variations across galactic disks \citep{Kreckel_2019_ApJ_887_80},
% the host galaxy of a fast radio burst source \citep{Xu_2022_Nature_609_685},
% the galactic azimuthal variations of oxygen abundance \citep{Ho_2017_ApJ_846_39},
% the mass-metallicity relation in galaxies \citep{BarreraBallesteros_2017_ApJ_844_80},
% the value-added catalog of galactic spectral energy distributions \citep{Chilingarian_2017_ApJS_228_14}
% the \ion{Ly}{$\alpha$} emitter at the epoch of reionisation \citep{Jung_2024_ApJ_967_73}.

\section{Conclusions}
% general message: be careful when including model error in the likelihood
IZI \citep{Blanc_2015_ApJ_798_99,Mingozzi_2020_A&A_636_A42} is a widely used Bayesian analysis tool for emission lines from \ion{H}{II} regions and star-forming galaxies.
Using photoionisation model grids, IZI returns the metallicity, ionisation parameter, and nebular emission-line colour excess estimates, expressed as \metal{}, \logq{}, and \EBV{}, respectively.
We noticed that IZI might underestimate or overestimate its parameters---\metal{}, \logq{}, and \EBV{}---judging from the results of \cite{Shinn_2020_MNRAS_499_1073}, and have studied the reliability of IZI estimates with the hope of fostering the appropriate use of IZI.
We created several mock datasets with reasonable input parameter values, ran IZI on the mock datasets, and checked how well IZI recovers the input parameter values.

Our test results show that IZI underestimates or overestimates parameters by approximately 1-$\sigma$ or more for the mock datasets designed for realistic observations when the model error is included (`mock\_snr\#\#\#\_[c-e]', see Figs.~\ref{fig:violinplots020}-\ref{fig:violinplots003}).
This issue arose even when we used all the emission lines provided by the photoionisation model grid (`mock\_snr010\_b' and `mock\_snr020\_b', see Fig.~\ref{fig:violinplots020} and Table \ref{tbl:config}).
The error of the predicted line flux given by the photoionisation model grids, that is, the model error, dominates the data error for all cases of SNR = 3, 10, and 20.
These results indicate that IZI can underestimate or overestimate the model parameters when applied to real observational data.
Therefore, we strongly recommend that IZI users run parameter recovery tests tailored to their data, such as in Figs.~\ref{fig:mock_snr010_c}-\ref{fig:mock_snr010_e}, before interpreting their IZI results.
To make this easier and encourage the appropriate use of IZI in the community, we have shared our Jupyter script used for the parameter recovery tests, including instructions (see the `Data Availability' section).

The cause of the underestimation or overestimation of IZI was the meeting of the following three factors: (i) the inclusion of the model error in the likelihood term, (ii) substantial size of the model error, and (iii) dependence of the model error on the model parameters.
We note that any parameter estimations where the likelihood includes the IZI-type model errors can underestimate or overestimate parameters in the same manner as IZI.
In this sense, NebulaBayes, another Bayesian analysis tool for the emission lines from photoionisation \citep{Thomas_2018_ApJ_856_89}, could suffer from biased parameter estimation, and we recommend that NebulaBayes users run parameter recovery tests. 
We also note that the \logq{} prior setting using the observed line ratio \SoS{} has two issues (the mismatch of line flux terms and violation of Bayes' theorem) and suggest a way to avoid these issues (Section \ref{sec:prior}).

We believe that IZI users will have more accuracy and reliability in the IZI parameter estimates if they use IZI in light of the findings of this study.

\section*{Acknowledgements}

% The Acknowledgements section is not numbered. Here you can thank helpful
% colleagues, acknowledge funding agencies, telescopes and facilities used etc.
The authors appreciate the comments from the anonymous referee, which significantly improved the manuscript.
J-HS is thankful to Min-Su Shin for valuable discussions on the Bayesian analysis.
J-HS and KO were supported by the Korea Astronomy and Space Science Institute under the R\&D programme (Project No. 2025-1-831-01), supervised by the Korea AeroSpace Administration.
RS acknowledges financial support from FONDECYT Regular 2023 project No. 1230441 and ANID - MILENIO NCN2024\_112 - MINGAL (Millennium Nucleus for GALaxies).
KO acknowledges support from the National Research Foundation of Korea (NRF) grant funded by the Korea government (MSIT) (RS-2025-00553982).

%%%%%%%%%%%%%%%%%%%%%%%%%%%%%%%%%%%%%%%%%%%%%%%%%%
\section*{Data Availability}
% The inclusion of a Data Availability Statement is a requirement for articles published in MNRAS. Data Availability Statements provide a standardised format for readers to understand the availability of data underlying the research results described in the article. The statement may refer to original data generated in the course of the study or to third-party data analysed in the article. The statement should describe and provide means of access, where possible, by linking to the data or providing the required accession numbers for the relevant databases or DOIs.
The Jupyter script for parameter recovery tests with IZI can be downloaded from \url{https://data.kasi.re.kr/vo/Stat_Reanal/}.



\bibliographystyle{mnras}
\bibliography{stat_reanal_2024a} % if your bibtex file is called example.bib

%\begin{thebibliography}{99}
%\end{thebibliography}

%%%%%%%%%%%%%%%%%%%%%%%%%%%%%%%%%%%%%%%%%%%%%%%%%%

%%%%%%%%%%%%%%%%% APPENDICES %%%%%%%%%%%%%%%%%%%%%

% \appendix

% \section{Some extra material}

% If you want to present additional material which would interrupt the flow of the main paper,
% it can be placed in an Appendix which appears after the list of references.

%%%%%%%%%%%%%%%%%%%%%%%%%%%%%%%%%%%%%%%%%%%%%%%%%%

% Don't change these lines
\bsp	% typesetting comment
\label{lastpage}
\end{document}

%% file: tbl/mock_test_setup.tex
mock\_snr100              &   100                     &  -$\infty$                &   8.8                     &   7.1                     &   0.1                     & \ion{C}{II} 1335, \ion{C}{IV} 1549, \ion{C}{III}] 1908, \ion{C}{II}] 2324, [\ion{O}{II}] 3726\\
                          &                           &                           &                           &                           &                           & [\ion{O}{II}] 3729, [\ion{Ne}{III}] 3869, \underline{[\ion{S}{II}] 4069}, H $\gamma$, \underline{[\ion{O}{III}] 4363}\\
                          &                           &                           &                           &                           &                           & \ion{He}{I} 4471, [\ion{O}{III}] 4959, [\ion{O}{III}] 5007, \ion{He}{I} 5016, [\ion{Ar}{III}] 5192\\
                          &                           &                           &                           &                           &                           & [\ion{N}{I}] 5198, \underline{[\ion{N}{II}] 5755}, \ion{He}{I} 5875, [\ion{O}{I}] 6300, \underline{[\ion{S}{III}] 6312}\\
                          &                           &                           &                           &                           &                           & [\ion{N}{II}] 6548, H $\alpha$, [\ion{N}{II}] 6584, \ion{He}{I} 6678, [\ion{S}{II}] 6717\\
                          &                           &                           &                           &                           &                           & [\ion{S}{II}] 6731, [\ion{Ar}{III}] 7136, \underline{[\ion{O}{II}] 7318}, [\ion{Ar}{III}] 7751, [\ion{S}{III}] 9068\\
                          &                           &                           &                           &                           &                           & [\ion{S}{III}] 9532      \\
mock\_snr010\_a           &    10                     &  -$\infty$                &   8.8                     &   7.1                     &   0.1                     & \ion{C}{II} 1335, \ion{C}{IV} 1549, \ion{C}{III}] 1908, \ion{C}{II}] 2324, [\ion{O}{II}] 3726\\
                          &                           &                           &                           &                           &                           & [\ion{O}{II}] 3729, [\ion{Ne}{III}] 3869, \underline{[\ion{S}{II}] 4069}, H $\gamma$, \underline{[\ion{O}{III}] 4363}\\
                          &                           &                           &                           &                           &                           & \ion{He}{I} 4471, [\ion{O}{III}] 4959, [\ion{O}{III}] 5007, \ion{He}{I} 5016, [\ion{Ar}{III}] 5192\\
                          &                           &                           &                           &                           &                           & [\ion{N}{I}] 5198, \underline{[\ion{N}{II}] 5755}, \ion{He}{I} 5875, [\ion{O}{I}] 6300, \underline{[\ion{S}{III}] 6312}\\
                          &                           &                           &                           &                           &                           & [\ion{N}{II}] 6548, H $\alpha$, [\ion{N}{II}] 6584, \ion{He}{I} 6678, [\ion{S}{II}] 6717\\
                          &                           &                           &                           &                           &                           & [\ion{S}{II}] 6731, [\ion{Ar}{III}] 7136, \underline{[\ion{O}{II}] 7318}, [\ion{Ar}{III}] 7751, [\ion{S}{III}] 9068\\
                          &                           &                           &                           &                           &                           & [\ion{S}{III}] 9532      \\
mock\_snr010\_b           &    10                     &   0.1$^\ddag$             &   8.8                     &   7.1                     &   0.1                     & \ion{C}{II} 1335, \ion{C}{IV} 1549, \ion{C}{III}] 1908, \ion{C}{II}] 2324, [\ion{O}{II}] 3726\\
                          &                           &                           &                           &                           &                           & [\ion{O}{II}] 3729, [\ion{Ne}{III}] 3869, \underline{[\ion{S}{II}] 4069}, H $\gamma$, \underline{[\ion{O}{III}] 4363}\\
                          &                           &                           &                           &                           &                           & \ion{He}{I} 4471, [\ion{O}{III}] 4959, [\ion{O}{III}] 5007, \ion{He}{I} 5016, [\ion{Ar}{III}] 5192\\
                          &                           &                           &                           &                           &                           & [\ion{N}{I}] 5198, \underline{[\ion{N}{II}] 5755}, \ion{He}{I} 5875, [\ion{O}{I}] 6300, \underline{[\ion{S}{III}] 6312}\\
                          &                           &                           &                           &                           &                           & [\ion{N}{II}] 6548, H $\alpha$, [\ion{N}{II}] 6584, \ion{He}{I} 6678, [\ion{S}{II}] 6717\\
                          &                           &                           &                           &                           &                           & [\ion{S}{II}] 6731, [\ion{Ar}{III}] 7136, \underline{[\ion{O}{II}] 7318}, [\ion{Ar}{III}] 7751, [\ion{S}{III}] 9068\\
                          &                           &                           &                           &                           &                           & [\ion{S}{III}] 9532      \\
mock\_snr010\_c           &    10                     &   0.1$^\ddag$             &   8.8                     &   7.1                     &   0.1                     & [\ion{O}{II}] 3726, [\ion{O}{II}] 3729, [\ion{O}{III}] 4959, [\ion{O}{III}] 5007, [\ion{N}{II}] 6548\\
                          &                           &                           &                           &                           &                           & H $\alpha$, [\ion{N}{II}] 6584, [\ion{S}{II}] 6717, [\ion{S}{II}] 6731\\
mock\_snr010\_d           &    10                     &   0.1                     &   8.8                     &   7.1                     &   0.1                     & \ion{C}{II}] 2324, [\ion{O}{II}] 3726, [\ion{O}{II}] 3729, [\ion{Ne}{III}] 3869, \underline{[\ion{S}{II}] 4069}\\
                          &                           &                           &                           &                           &                           & H $\gamma$, \underline{[\ion{O}{III}] 4363}, \ion{He}{I} 4471, [\ion{O}{III}] 4959, [\ion{O}{III}] 5007\\
mock\_snr010\_e           &    10                     &   0.1                     &   8.8                     &   7.1                     &   0.1                     & [\ion{O}{II}] 3726, [\ion{O}{II}] 3729, [\ion{O}{III}] 4959, [\ion{O}{III}] 5007\\
mock\_snr020\_a$^\dag$    &    20                     &  -$\infty$                &   8.8                     &   7.1                     &   0.1                     & same as mock\_snr010\_a  \\
mock\_snr020\_b$^\dag$    &    20                     &   0.1$^\ddag$             &   8.8                     &   7.1                     &   0.1                     & same as mock\_snr010\_b  \\
mock\_snr020\_c$^\dag$    &    20                     &   0.1$^\ddag$             &   8.8                     &   7.1                     &   0.1                     & same as mock\_snr010\_c  \\
mock\_snr020\_d$^\dag$    &    20                     &   0.1                     &   8.8                     &   7.1                     &   0.1                     & same as mock\_snr010\_d  \\
mock\_snr020\_e$^\dag$    &    20                     &   0.1                     &   8.8                     &   7.1                     &   0.1                     & same as mock\_snr010\_e  \\
mock\_snr003\_a$^\dag$    &     3                     &  -$\infty$                &   8.8                     &   7.1                     &   0.1                     & same as mock\_snr010\_a  \\
mock\_snr003\_b$^\dag$    &     3                     &   0.1$^\ddag$             &   8.8                     &   7.1                     &   0.1                     & same as mock\_snr010\_b  \\
mock\_snr003\_c$^\dag$    &     3                     &   0.1$^\ddag$             &   8.8                     &   7.1                     &   0.1                     & same as mock\_snr010\_c  \\
mock\_snr003\_d$^\dag$    &     3                     &   0.1                     &   8.8                     &   7.1                     &   0.1                     & same as mock\_snr010\_d  \\
mock\_snr003\_e$^\dag$    &     3                     &   0.1                     &   8.8                     &   7.1                     &   0.1                     & same as mock\_snr010\_e  